\documentclass[12pt]{article}
\usepackage{color}
\usepackage{amssymb,amsmath}
\usepackage[dvipdfmx]{graphicx}

\newcommand{\bea}{\begin{eqnarray}}
\newcommand{\eea}{\end{eqnarray}}
\begin{document}
\setlength{\baselineskip}{0.7cm}
\setlength{\baselineskip}{0.7cm}
\begin{titlepage} 
%%%%% PREPRINT NUMBERS %%%%%%
\begin{flushright}
OCU-PHYS 474 \\
YGHP-18-01 
\end{flushright}
\vspace*{10mm}
%%%%%%%%%%%%%%%%%%% TITLE %%%%%%%%%%%%%%%%%%
\begin{center}{\Large\bf $SU(2)_L$ Doublet Vector Dark Matter \\
\vspace*{2mm}
from  Gauge-Higgs Unification}
\end{center}
%%%%%%%%%%%%%%%% AUTHORS %%%%%%%%%%%%%%%%%%%%%%%
\vspace*{10mm}
\begin{center}
{\large Nobuhito Maru}$^{a}$, %\footnote{E-mail: nmaru@sci.osaka-cu.ac.jp} 
{\large Nobuchika Okada}$^{b}$, %\footnote{E-mail: okadan@ua.edu}
and 
{\large Satomi Okada}$^{c}$%\footnote{E-mail: s15e102d@st.yamagata-u.ac.jp}
\end{center}
%%%%%%%%%%%%%%%%%%%%%%% AFFILIATION %%%%%%%%%%%%
\vspace*{0.2cm}
\begin{center}
%\small
${}^{a}${\it 
Department of Mathematics and Physics, Osaka City University, \\ 
Osaka 558-8585, Japan}
\\[0.2cm]
${}^{b}${\it Department of Physics and Astronomy, University of Alabama, \\
Tuscaloosa, Alabama 35487, USA} 
\\[0.2cm]
${}^{c}${\it Graduate School of Science and Engineering, Yamagata University, \\
Yamagata 990-8560, Japan} 
%%%%%
\end{center}
%%%%%%%%%%%%%%%%%% ABSTRACT %%%%%%%%%%%%%%%
\vspace*{1cm}
\begin{abstract} 

A new vector dark matter (DM) scenario in the context of the gauge-Higgs unification (GHU) is proposed. 
The DM particle is identified with an electric-charge neutral component in an $SU(2)_L$ doublet vector field 
  with the same quantum number as the Standard Model Higgs doublet. 
Since such an $SU(2)_L$ doublet vector field is incorporated in any models of the GHU scenario,  
  it is always a primary and model-independent candidate for the DM in the scenario. 
The observed relic density is reproduced through a DM pair annihilations into the weak gauge bosons 
  with a TeV-scale DM mass, which is nothing but the compactification scale of extra-dimensions. 
Due to the higher-dimensional gauge structure of the GHU scenario, 
  a pair of the DM particles has no direct coupling with a single $Z$-boson/Higgs boson, 
  so that the DM particle evades the severe constraint from the current direct DM search experiments.  

\end{abstract}
\end{titlepage}
%%%%%%%%%%

The existence of the dark matter (DM) in our universe is undoubtedly believed 
  from the various cosmological observations, and the DM particle is one of the key ingredients
  in exploring physics beyond the Standard Model (SM). 
Identification with the DM particle is still one of the unsolved problems in particle physics and cosmology.  
In this paper, we consider the DM physics in the gauge-Higgs unification (GHU) scenario \cite{GH}. 
This scenario can provide us with a solution to the gauge hierarchy problem 
  without invoking supersymmetry, 
  where the SM Higgs doublet is identified with an extra spatial component of the gauge field 
  in higher dimensions. 
It is remarkable that irrespective of the non-renormalizability of the theory,  
  the GHU scenario predicts various finite physical observables,  
  such as Higgs potential~\cite{1loopmass, 2loop},  $H\to gg, \gamma\gamma$~\cite{MO, Maru, diphoton}, 
  the anomalous magnetic moment $g-2$~\cite{g-2} and the electric dipole moment~\cite{EDM}, 
  thanks to the higher dimensional gauge symmetry. 

In this paper, we propose a new possibility that an electric-charge neutral component in an $SU(2)_L$ doublet gauge field 
  is identified with the DM particle, where $SU(2)_L$ is the gauge group of the weak interaction in the SM. 
The vector DM particle as the lightest Kaluza-Klein (KK) mode of the photon 
  in the Universal Extra Dimension (UED) model \cite{UED_DM} 
  or a $T$-odd partner of the photon in the Littlest Higgs model with $T$-parity \cite{LHT_DM}
  has been extensively studied. 
However, as far as we know, a vector DM in the $SU(2)_L$ doublet, 
  in general, non-adjoint representation of $SU(2)_L$ has not been commonly studied.\footnote{
See Ref.~\cite{331} for a model proposal.  
Collider physics for the $SU(2)_L$ doublet vector field has been investigated in Ref.~\cite{Agashe}
}   
Although we may introduce such a particle into the SM,  
  we notice that the $SU(2)_L$ doublet vector field is automatically incorporated in any models of the GHU.

In the GHU scenario, the SM gauge group is extended and embedded into larger gauge groups 
  in higher-dimensional space-time. 
The SM $SU(2)_L$ doublet Higgs is identified with  a spatial component of the higher dimensional gauge field. 
Because of this GHU structure, the $SU(2)_L$ doublet gauge field 
  with the same quantum number as the SM Higgs doublet 
  is automatically contained in a coset space of the extended gauge group 
  as a ``gauge-Higgs partner'' of the Higgs boson. 
The existence of the $SU(2)_L$ doublet gauge field, as a gauge-Higgs partner 
  of the SM Higgs doublet, is therefore a model-independent feature of the GHU scenario.  
In order to reproduce the SM gauge group, a non-trivial orbifold boundary condition 
  is imposed in the scenario and, as a result, a vector field of the gauge-Higgs partner 
  to the SM Higgs boson emerges as an odd particle under a $Z_2$-symmetry.  
This is a DM candidate of the scenario.  
Our main purpose of this paper is to show the basic structure of the GHU scenario 
  based on a simple $SU(3)$ model in 5-dimensional (5D) space-time and propose a new vector DM scenario. 
One of our remarkable results is that due to the higher-dimensional gauge structure of the GHU scenario, 
  a coupling between a pair of the DM particles and a single $Z$-boson/Higgs boson is absent, 
  and as a result, the DM particle evades the severe constraint from the current direct DM detection experiments. 
Through its pair annihilations into the weak gauge bosons, the observed DM relic density 
  is reproduced with a TeV-scale DM mass, which is nothing but the compactification scale of the 5th dimension. 
For related DM scenarios in the context of GHU, see Ref.~\cite{GHDM}.

Let us consider a simple GHU model based on the gauge group $SU(3)$ 
  in a flat 5D Minkowski space with the 5th dimension 
  compactified on an orbifold $S^1/Z_2$ with radius $R$ of $S^1$. 
Since reproducing the realistic Yukawa couplings for the SM fermions is not our main scope, 
  we simply follow Ref.~\cite{CCP} for our setup on bulk fermions corresponding to the SM fermions. 
As is well-known, the simple $SU(3)$ model cannot provide a realistic weak mixing angle. 
Following Ref.~\cite{SSS}, we introduce brane localized kinetic terms for the SM gauge bosons 
  and adjust them so as to reproduce the realistic weak mixing angle. 
The brane localized kinetic terms also plays an important role in our DM scenario 
 as will be discussed later.

The boundary conditions should be appropriately assigned to reproduce the SM fields as the zero modes. 
While a boundary condition corresponding to $S^1$ is taken to be periodic for all of the bulk SM fields, 
  the $Z_2$ parity is assigned for gauge fields and fermions in the representation ${\cal R}$ 
  by using the parity matrix $P={\rm diag}(-,-,+)$ as
\bea
 A_\mu (-y) = P^\dag A_\mu(y) P, \quad A_y(-y) =- P^\dag A_y(y) P,  \quad 
\psi(-y) = {\cal R}(P) \gamma^5 \psi(y) ,
\label{parity}
\eea 
 where the subscripts $\mu$ ($y$) denotes the four (the fifth) dimensional component. 
The parity assignment breaks the $SU(3)$ gauge symmetry down to  the SM gauge group of 
  $SU(2)_L \times U(1)_Y$. 
Off-diagonal blocks in $A_y$, which correspond to an $SU(2)_L$ doublet, 
   have zero modes due to the overall sign in Eq.~(\ref{parity}). 
In fact, these are identified with the SM Higgs doublet ($H=(H^+ \;  H^0)^T$): 
\bea
%A_y^{(0)} = \frac{1}{\sqrt{2}}
%\left(
%\begin{array}{cc}
%0 & H \\
%H^\dag & 0 \\
%\end{array}
%\right). 
%
A_y^{(0)}  = \frac{1}{\sqrt{2}}
\left( 
\begin{array}{ccc}
0 & 0 & H^+ \\
0 & 0 & H^0 \\
H^- & H^{0*} & 0 \\
\end{array}
\right). 
\label{Mat_H}
\eea
The KK modes of $A_y$ are eaten by KK modes of the SM gauge bosons 
  and become their longitudinal degrees of freedom like the ordinary Higgs mechanism.

The 5D Lagrangian relevant to our DM physics is given by 
\bea
{\cal L}_{{\rm DM}} &=& 
- \frac{1}{2} \, {\rm Tr} \left[ F_{MN} F^{MN} \right] 
\nonumber \\
 &&- \left(
 \frac{c_L}{2} \, {\rm Tr} \left[W_{\mu\nu} W^{\mu\nu} \right]  
 +\frac{c_Y}{4} \, {\rm Tr} \left[B_{\mu\nu} B^{\mu\nu} \right] 
% +\frac{c_X}{2} \, \left(X_{\mu\nu} \right)^\dagger X^{\mu\nu}  
  \right)
(
 \delta(y) + \delta(y-\pi R)
 ) 
\label{DMLagrangian}
\eea
where $M,N$ and $\mu, \nu$ are 5D and 4D indices respectively. 
The first term in the right-hand side is a 5D gauge kinetic term of the $SU(3)$ gauge field, 
  while the remaining terms are the brane localized kinetic terms at orbifold fixed points $y=0$ and $\pi R$.   
Since the bulk $SU(3)$ gauge symmetry is explicitly broken to $SU(2)_L \times U(1)_Y$ at the fixed points, 
   we have introduced the brane kinetic terms 
   for the $SU(2)_L$ and the $U(1)_Y$ gauge bosons, 
   where $W_{\mu \nu}$ and $B_{\mu \nu}$ are their field strengths, respectively. 
Adjusting suitable values for the parameters, $c_L$ and $c_Y$, 
   the realistic weak mixing angle can be reproduced in 4D effective theory \cite{SSS}. 
%Here, we have also introduced the brane localized kinetic term 
%   for a $SU(2)_L$ doublet gauge boson (whose filed strength is denoted as $X_{\mu \nu})$) with a parameter $c_X$, 
%   which includes the DM candidate as we will discuss in the following. 
We here emphasize that the brane localized kinetic terms for each gauge boson have been set 
   to be exactly the same at $y=0$ and $\pi R$ to preserve the KK-parity \cite{CTW}.

Let us now look at the terms in the Lagrangian involving the DM candidate. 
Following the parity in Eq.~(\ref{parity}), 
  the SM gauge fields and the Higgs doublet are embedded in $A_\mu$  and $A_y$, respectively, 
  as zero-modes.
In the following calculations, it may be convenient to express them explicitly in the matrix form: 
\bea
A_\mu^{(0)} = \frac{1}{2}
\left( 
\begin{array}{ccc}
W_\mu^3 + \frac{1}{\sqrt{3}} B_\mu & \sqrt{2} W_\mu^+ & 0 \\
\sqrt{2} W_\mu^- & -W_\mu^3 + \frac{1}{\sqrt{3}} B_\mu & 0 \\
0 & 0 & -\frac{2}{\sqrt{3}}B_\mu \\
\end{array}
\right), 
\label{Mat_A}
\eea 
for the SM gauge fields\footnote{
In the presence of the brane localized kinetic terms, 
  the normalization of the SM gauge fields, $W_\mu^{\pm}$, $W_\mu^3$ and $B_\mu$, 
  are not yet canonical. 
Since the coupling manner of the DM particle with the SM particles is independent of 
  the normalization, we proceed our calculations with the present normalization. 
} and Eq.~(\ref{Mat_H}) for the Higgs doublet. 
The DM candidate is embedded  in $A_\mu$ 
  as the first KK mode of the gauge-Higgs partner of the Higgs doublet. 
Its matrix form is given by 
\bea
A_\mu^{(1)} = \frac{1}{\sqrt{2}}
\left( 
\begin{array}{ccc}
0 & 0 & X_\mu^1 \\
0 & 0 & X_\mu^2 \\
X_\mu^{1*} & X_\mu^{2*} & 0 \\
\end{array}
\right). 
\label{Mat_DM}
\eea
Here, $X_\mu^2$ is an electric-charge neutral component. 
In the following, it turns out that the DM candidate is the real part of $X_\mu^2$, 
  which is nothing but the gauge-Higgs partner of the SM Higgs boson.

The mass spectrum of the $SU(2)_L$ doublet vector 
  in 4D effective theory is calculated from 
  a part of the 5D gauge kinetic term ${\rm Tr}(F_{\mu y}F^{\mu y})$.
After the electroweak symmetry breaking by a vacuum expectation value (VEV) of $\langle H^0 \rangle=v/\sqrt{2}$, 
  we have the mass spectrum for the electric-charge neutral component $X_\mu^2$: 
\bea
\int dy \, {\cal L}_{{\rm DM}} &\supset & -\int dy \,  {\rm Tr}\left[F_{\mu y}F^{\mu y}\right] \nonumber \\
 &\supset& 
 -\frac{g^2v^2}{8} \eta^{\mu \nu} (X^2_\mu-X^{2*}_\mu)(X^2_\nu-X^{2*}_\nu)
  +\frac{1}{2R^2}  \eta^{\mu \nu} (X^{2*}_\mu X^2_\nu) \nonumber \\ 
 &=&\frac{1}{2}  \left(\frac{1}{R} \right)^2  \eta_{\mu \nu} \left( X^\mu_{{\rm DM}}  X^\nu_{\rm DM} \right)
 +\frac{1}{2} \left( \left( \frac{1}{R} \right)^2 + 4 m_W^2 \right) X^\mu X_\mu ,
\eea
where $X_{{\rm DM} \mu}$ and $X_\mu$ are defined as
\bea
X_{{\rm DM} \mu } \equiv \frac{1}{\sqrt{2}}(X^2_\mu + X^{2*}_\mu), \quad 
X_{\mu} \equiv -\frac{i}{\sqrt{2}}(X^2_\mu - X^{2*}_\mu). 
\label{Def_DM}
\eea
The distinctive feature can be seen as follows. 
The real part of $X_\mu^2$ (denoted as $X_{\rm DM}^\mu$ hereafter) receives no electroweak symmetry breaking mass 
 proportional to the Higgs doublet VEV, while its imaginary part receives it: 
 $m^2_{{\rm DM}} = \left(\frac{1}{R} \right)^2$, $m_X^2 = \left(\frac{1}{R} \right)^2 + 4m_W^2$. 
As a result, $X_{\rm DM}^\mu$ is the lightest mode and hence the DM candidate. 
Note that since the DM particle receives no electroweak symmetry breaking mass, 
  its pair has no coupling with a single Higgs boson.

One may think that the 1st KK mode of the photon has the mass $1/R$ 
  and can be the DM candidate as in the UED models. 
In order to see which of 1st KK gauge bosons is the lightest one, 
  we must take into account the effects of the brane localized kinetic terms. 
Such effects on the KK-mode mass spectrum have been analyzed in Ref.~\cite{CTW}. 
%When the overall signs for the brane localized kinetic terms 
%  are opposite to those in the setup of  \cite{CTW}, 
%  we can easily find that the KK mode masses become larger than $n/R$.  
When we take $c_L, c_Y < 0$ in Eq.~(\ref{DMLagrangian}), 
  we can find that the $n$-th KK mode mass is larger than $n/R$.    
In fact, let us consider the eigenvalue equation of Eq.~(33) in Ref.~\cite{CTW} 
  with a common coefficients for the brane localized kinetic terms ($r_a=r_b$).  
For $|r_a m_n| \ll 1$, the eigenvalue equation is approximately given by 
\bea
\tan (m_n \pi R) \simeq - r_a, 
\eea
which means that the $n$-th KK mode mass is larger than $n/R$ for $r_a < 0$.\footnote{
$|r_a|$ cannot be very large in order to keep 4D effective gauge coupling squared positive. 
} 
Hence, the lightest KK gauge boson is found as $X_{\rm DM}^\mu$.

One may also think that the 1st KK modes of the SM fermions have the mass of $1/R$. 
This is not true since the $Z_2$-odd bulk mass terms are introduced 
  to obtain the realistic Yukawa couplings except for top quark \cite{CCP} 
  and the masses for the 1st KK modes become heavier by the bulk mass ($M$) contributions 
  such as $\sqrt{M^2 + \left(\frac{1}{R} \pm m_f \right)^2}$,  
  where $m_f$ is a fermion mass generated by the Higgs doublet VEV \cite{HKLT}.  
One should note that the $Z_2$-odd bulk mass terms preserve the KK-parity \cite{splitUED}. 
For top quark, the bulk mass must be zero to reproduce the top quark mass. 
Therefore, one might conclude that the lightest KK particle is the 1st KK top quark 
  with its mass eigenvalue $1/R  - m_t$. 
However, for more precise evaluation for the KK-mode mass spectrum, 
  we need to take into account the 1-loop quantum corrections to the 1st KK top quark mass. 
As has been calculated in Ref.~\cite{CMS}, this corrections are dominated by QCD effects 
  with a logarithmic divergence, which makes the resultant 1st KK top quark mass 
  sizably larger than $1/R$.   
On the other hand, quantum corrections to the KK mode gauge boson masses are found to be finite 
  because of the original 5D gauge invariance \cite{CMS}. 
Since the 1st KK $SU(2)_L$ gauge boson $X_\mu$ receives such finite corrections proportional to 
  $SU(2)_L$ and $U(1)_Y $  gauge coupling squareds, 
  we find that the 1st KK top quark is heavier than the 1st KK mode of $X_\mu$. 
Therefore,  $X_{\rm DM}^\mu$ is a unique DM candidate.
%One may also think that the 1st KK mode of the photon has the mass $1/R$ 
%  and can be the DM candidate as in the UED models. 
%In order to see which of 1st KK gauge bosons is the lightest one, 
%  we must take into account the effects of the brane localized kinetic terms. 
%Such effects on the KK-mode mass spectrum have been analyzed in Ref.~\cite{CTW}. 
%It has been found that the 1st KK mode masses become smaller than $1/R$ 
%   as the coefficients of the brane localized kinetic terms are larger. 
%Hence, if we take parameters as $c_X > c_{W,Y}$, 
%  the lightest KK gauge boson is found as $X_{\rm DM}^\mu$. 
%As we have mentioned above, 
%   our setup maintains the KK-parity even in the presence of the brane localized kinetic terms, 
%   and therefore $X_{\rm DM}^\mu$ is a unique DM candidate.
  %In order to realize the right value for the weak mixing angle, 
%  we have introduced the brane kinetic terms for the bulk SM gauge boson at $y=0$, 
%  which explicitly violate the KK-parity. 
%As a result, the KK-mode functions of the SM gauge bosons become 
%  non-trivial \cite{CTW} and the KK-mode SM gauge bosons turn to be unstable 
%  through interactions with a pair of SM fermions. 
%Therefore, $X_{\rm DM}^\mu$ is a unique DM candidate. 

We have found that a pair of the vector DM ($X_{\rm DM}^\mu$) has no coupling with a single Higgs boson.
Here we show that a triple vector coupling among a DM pair and $Z$-boson is also absent. 
If it exists, such a triple vector coupling is included in 
 the 3-point self interaction of the 5D $SU(3)$ gauge boson: 
\bea  
 {\cal L}_{5 D} \supset  -\frac{1}{2} \, {\rm Tr} \left[ F_{MN} F^{MN} \right] 
  \supset ig {\rm Tr}[(\partial_\mu A_\nu - \partial_\nu A_\mu)[A^\mu, A^\nu]]. 
\eea
By using the explicit forms in Eqs.~(\ref{Mat_A}) and (\ref{Mat_DM}) 
  in terms of Eq.~(\ref{Def_DM}), we find
\bea
&& 
ig \,  {\rm Tr}[(\partial_\mu A_\nu - \partial_\nu A_\mu)[A^\mu, A^\nu]]  \nonumber \\
&\supset& 
\frac{ig}{4} \left[
-2 \partial_\mu \left(W_\nu^3 + \frac{1}{\sqrt{3}}B_\nu \right)
(-X^\mu_{{\rm DM}}X^\nu + X^\nu_{{\rm DM}}X^\mu) 
\right. \nonumber \\ 
&& \left. + 2 \partial_\mu X_{{\rm DM} \nu} 
\left\{-X^\mu(-W^{3\nu} + \sqrt{3}B^\nu) + X^\nu(-W^{3\mu} + \sqrt{3}B^\mu) \right\}
\right. \nonumber \\
&& \left. 
+2 \partial_\mu X_\nu 
\left\{X^\mu_{\rm DM}(-W^{3\nu} + \sqrt{3}B^\nu) - X^\nu_{{\rm DM}}(-W^{3\mu} + \sqrt{3}B^\mu) \right\}
\right]. 
\eea
Now we see that the triple vector coupling among a DM pair and $Z$-boson 
  (which is a liner combination of $W_\mu^3$ and $B_\mu$) is absent in the model.

Finally, we discuss the relic density of the vector DM particle. 
All couplings of the vector DM with the SM particles are derived from the original 5D Lagrangian, 
  ${\cal L}_{5 D} \supset  -\frac{1}{2} \, {\rm Tr} \left[ F_{MN} F^{MN} \right]$. 
Because of the vector nature, there are so many interaction terms in 4D effective Lagrangian,
  and the complete analysis for the relic density is very tedious. 
Thus, we here perform a rough estimate of the DM relic density.  
In 4D effective theory, we can express the kinetic term for the DM $SU(2)_L$ doublet as 
\bea 
   {\cal L}_{4D} = - \frac{1}{2}  (D_\mu \chi_\nu -D_\nu \chi_\mu)^\dagger  (D^\mu \chi^\nu -D^\nu \chi^\mu),  
\eea
where $\chi_\mu = (X_\mu^1 \; X_\mu^2)^T$ is the DM $SU(2)_L$ doublet, 
  and  $D_\mu$ is the covariant derivative, which is the same as the covariant derivative 
  for the SM Higgs doublet. 
For our estimate of the DM relic density, let us use the 4-point interactions 
  among the DM particles and the $W$-bosons, which can be extracted as
\bea 
   {\cal L}_{4D} 
&=& 
    - \frac{1}{2}  (D_\mu \chi_\nu -D_\nu \chi_\mu)^\dagger  (D^\mu \chi^\nu -D^\nu \chi^\mu) \nonumber \\
&\supset & 
 - g^2 \left( \eta^{\mu \nu} \eta^{\rho \sigma} - \eta^{\mu \rho} \eta^{\nu \sigma} \right) 
   X_{{\rm DM} \mu} X_{{\rm DM} \nu} W^{+}_\rho W^{-}_\sigma .
\eea  
Using this interactions, we calculate the DM pair annihilation cross section in the non-relativistic limit as 
\bea 
  \sigma v_{\rm rel} = \frac{5 g^4}{48 \pi m_{\rm DM}}, 
\eea
where $v_{\rm rel}$ is a relative velocity of two annihilating DM particles, and $m_{\rm DM}$ is a DM mass. 
It is well known that the annihilation cross section of $\sigma v_{\rm rel}=1$ pb 
  is required to reproduce the observed DM relic abundance \cite{KT},  
  which is achieved by the DM mass of $m_{\rm DM} \simeq 1.5$ TeV. 
Here, we have used $g\simeq 0.65$ for the $SU(2)_L$ gauge coupling.

In summary, we have proposed a new vector DM scenario in the context of the GHU, 
  where the DM particle is included in the 1st KK mode of the $SU(2)_L$ doublet vector field.  
Because of the orbifold boundary condition, the 1st KK mode of the $SU(2)_L$ doublet vector field 
  is odd under the $Z_2$-parity in 4D effective theory and hence the lightest component is stable. 
We have found that the real component of electric-charge neutral component in the $SU(2)_L$ doublet vector, 
  which is the gauge-Higgs partner of the Higgs boson, 
  is the lightest and hence the DM candidate. 
Since the $SU(2)_L$ doublet vector is incorporated in any GHU models 
  as the gauge-Higgs partner of the Higgs doublet, 
  our proposal is model-independent and the electric charge-neutral component of  
   the $SU(2)_L$ doublet vector is a primary DM candidate in the GHU scenario. 
Based on a simple $SU(3)$ GHU model in the flat 5D Minkowski space, 
   we  have shown the basic structure of the DM scenario. 
The DM mass is determined by the compactified radius with no contribution 
   from the Higgs doublet VEV.  
Because of the higher-dimensional gauge structure of the GHU model, 
   a coupling among a DM particle pair and a single $Z$-boson/Higgs boson 
   is absent in the model. 
Thus, the vector DM particle evades the severe constraint from the current direct DM detection experiments. 
The observed relic abundance of the DM particle is obtained through its pair annihilation processes 
  into the weak gauge bosons for a TeV-scale DM mass. 
This DM mass, which is nothing but the compactification scale of the GHU model, 
  is the natural scale for the GHU scenario to provide a solution to the gauge hierarchy problem.

%%%%%%%%%%%%%%%
\section*{Acknowledgments}
%%%%%%%%%%%%%%%
This work is supported in part by JSPS KAKENHI Grant Number JP17K05420 (N.M.) 
  and the DOE Grant Number DE-SC0012447 (N.O.).

%%%%%%%%%%%

\end{document}